\documentclass[%
 aip,
 amsmath,amssymb,
reprint,%
]{revtex4-1}

\usepackage{graphicx}
\usepackage{subfigure}
\usepackage{dcolumn}
\usepackage{bm}
\usepackage{color}
\usepackage[colorlinks=true,citecolor=blue,linkcolor=blue]{hyperref}

\begin{document}
\newcommand{\SR}[1]{\noindent \color{red}  #1 \normalcolor}
\newcommand{\PG}[1]{\noindent \color{blue}  #1 \normalcolor}
\newcommand{\Fder}{\mathbf{F}}
\newcommand{\n}{\mathbf{n}}
\newcommand{\B}{\mathbf{B}}
\newcommand{\I}{\mathbf{I}}
\newcommand{\kk}{\mathbf{k}}
\newcommand{\e}{\mathbf{e}}
\newcommand{\dd}{\mathbf{d}}

\preprint{AIP/123-QED}

\title[Explicit dispersion relations for elastic waves in extremely deformed soft matter]{Explicit dispersion relations for elastic waves in extremely deformed soft matter with application to nearly incompressible and auxetic materials}

\author{Pavel Galich}
\author{Stephan Rudykh}\email{rudykh@technion.ac.il}
\affiliation{Faculty of Aerospace Engineering, Technion - Israel Institute of Technology, Haifa 32000, Israel}

\date{\today}

\begin{abstract}
We analyze the propagation of elastic waves in soft materials subjected to finite deformations. We derive explicit dispersion relations, and apply these results to study elastic wave propagation in (i) nearly incompressible materials such as biological tissues and polymers, and (ii) negative Poisson's ratio or auxetic materials. We find that for nearly incompressible materials transverse wave velocities exhibit strong dependence on direction of propagation and initial strain state, whereas the longitudinal component is not affected significantly until extreme levels of deformations are attained.
For highly compressible materials, we show that both pressure and shear wave velocities depend strongly on initial deformation and direction of propagation. When compression is applied, longitudinal wave velocity decreases in positive bulk modulus materials, and increases for negative bulk modulus materials; this is regardless the direction of wave prorogation. We demonstrate that finite deformations influence elastic wave propagation through combinations of induced effective compressibility and stiffness.

\end{abstract}

\maketitle

The propagation of elastic waves has been investigated intensively\cite{Biot1940,Truesdell1965,Norris1983,Ogden1984,Kushwahaetal93prl,Ignatovich2009,OgdenSighn2011jmms,Shamsetal2011,Parnell2012,DestradeOgden13jam,Hussein2014} because the understanding of the phenomenon is vital for a large variety of applications from non-invasive materials testing and medical imaging for health care to petroleum exploration. Recently, the field of acoustic or phononic metamaterials has attracted a considerable attention. The peculiarity of these metamaterials originates in their microstructure, which can be tailored to give rise to various effects such as local resonances\cite{Wang&etal2014prl}, band-gaps\cite{Hussein2014} and cloaking\cite{Fang2011prl}. 
Furthermore, \emph{soft} metamaterials, due to their capability to sustain large deformations, open promising opportunities of manipulating acoustic characteristics via deformation \cite{BertoldiBoyce2008prb,RudykhBoyce2014prl,Li13etal}. 

In this work, we derive explicit dispersion relations for \emph{finitely deformed} materials by means of the superimposed on large deformations Bloch wave analysis frequently used for investigation of elastic wave propagation \cite{Kushwahaetal93prl,BertoldiBoyce2008prb,RudykhBoyce2014prl,Lydon&etal2014prl, Hussein2014}. The availability of the explicit dispersion relations is important for designing \emph{mechanotunable} acoustic metamaterials. Moreover, the information may greatly benefit non-invasive medical diagnostic techniques by providing the important information on dependence of elastic wave propagation on pre-stress/pre-strain conditions, which is a common state of biological tissues. By application of the derived explicit expressions, we show the role of deformation in wave propagation in soft media undergoing finite deformations. Moreover, we extent the analysis to a class of exotic metamaterials characterize by negative Poisson's ratio (NPR) behaviour. Examples of the NPR materials, also known as \emph{auxetics}, include living bone tissue\cite{WILLIAMS1982}, skin\cite{Evans2000}, blood vessels\cite{Caddock1995}, certain rocks and minerals\cite{Evans2000}, and artificial materials \cite{bertoldi2013}. 
 As we shall show, elastic wave propagation in these materials is significantly affected by deformation.

To analyse the finitely deformed state, we introduce the deformation gradient $\Fder (\mathbf{X},t)=\nabla_{\mathbf{X}}\otimes\mathbf{x(X},t)$, where $\mathbf{X}$ and $\mathbf{x}$ are position vectors in the reference and current configurations, respectively. To take into account finite deformation non-linear effects as well as material non-linearity, we analyse the wave propagation in terms of infinitesimal plane waves \emph{superimposed} on a finitely deformed state\cite{Truesdell1965,Ogden1984}. To account for the stiffening effects (due to, for example, finite extensibility of polymer chains, or due to collective straightening of collagen fibers in biological tissues) in finitely deformed media, we make use of the strain-energy density function corresponding to an approximation of Arruda-Boyce model\cite{Arruda1993}, namely the Gent model\cite{gent1996}
\begin{equation}\label{Gent}
	\begin{split}
		\psi(\Fder)=&-\frac{\mu J_m}{2}\ln\ \left(1-\frac{I_1-3}{J_m}\right)-\mu\ln\ J\\
		&+\left( \frac{K}{2}-\frac{\mu}{3}-\frac{\mu}{J_m} \right)(J-1)^2,
	\end{split}	
\end{equation}
where $\mu$ is the initial shear modulus, $K$ is the bulk modulus, $I_1=\text{tr}\  \B$ is the first invariant of the left Cauchy-Green tensor $\B=\Fder\Fder^T$ and $J=\text{det}\ \Fder$. The model neatly covers the stiffening of the material with the deformation; as the deformation attends the level of $I_1=3+J_m$, the energy function becomes unbounded and a dramatic increase in stress occurs. Consequently, $J_m$ is locking parameter. Clearly, when $J_m\rightarrow\infty$, the strain-energy function $\eqref{Gent}$ reduces to the neo-Hookean one, namely
\begin{equation}\label{neoH}
\psi(\Fder)=\frac \mu{2} (I_1-3)-\mu\ln\ J+\left( K/2-\mu/3 \right)(J-1)^2.
\end{equation}

Acoustic tensor for finitely deformed neo-Hookean material $\eqref{neoH}$ takes the form
\begin{equation}\label{acten}
\mathbf{Q(n)}=a_1\n\otimes\n + a_2(\I-\n\otimes\n),
\end{equation}
where the unit vector $\n$  defines the direction of propagation of the wave; $\n\otimes\n$ is the projection on the direction $\n$;  $(\I-\n\otimes\n)$ is the projection on the plane  normal to $\n$; $a_1=(K-2\mu/3)J+\mu J^{-1}(1+\n\cdot\B\cdot\n)$ and $a_2=\mu J^{-1}(\n\cdot\B\cdot\n)$. Consequently, there always exist one longitudinal (pressure) and two transverse (shear) waves for any direction of propagation $\n$. Phase velocities of these waves can be calculated as
\begin{equation}
 c_{pw}=\sqrt{a_1/\rho}\quad  \text{and}\quad  c_{sw}=\sqrt{a_2/\rho},\label{phv}
  \end{equation}
  where $\rho$ is the material density in the deformed state. In the limit of small deformations the results reduce to the well known relations $c_{pw}=\sqrt{(K+4\mu/3)/\rho}$ and $c_{sw}=\sqrt{\mu/\rho}$.

We present the results in the generic form where possible, however, the illustrative examples are given for uniaxial tension
\begin{equation}
 \Fder=\lambda\e_1\otimes\e_1+\tilde{\lambda}(\e_2\otimes\e_2+\e_3\otimes\e_3),\label{ut}
 \end{equation}
 where $\lambda$ is the applied stretch ratio and $\tilde{\lambda}=\tilde{\lambda}(\lambda,K/\mu)$ is defined through $\lambda$ and compressibility of the material.
Note that the compressibility of the material is defined by ratio $K/\mu$. In the limit of linear elasticity the elastic moduli are related through
 \begin{equation}\label{elm}
 \frac{K}{\mu}=\frac{2(1+\nu)}{3(1-2\nu)},
 \end{equation}
 where $\nu$ is Poisson's ratio. Thus, $-1/3<K/\mu<\infty$ with $\mu>0$. Note that for $-1/3<K/\mu<0$ the material is stable only if constrained\cite{ryzhak93jmps,wanglakes2005,Gasparetal2003jap}. Besides, it follows from $\eqref{elm}$ that matter exhibits auxetic behaviour when $-1/3<K/\mu<2/3$.

\begin{figure}[h]
  \includegraphics[scale=0.3]{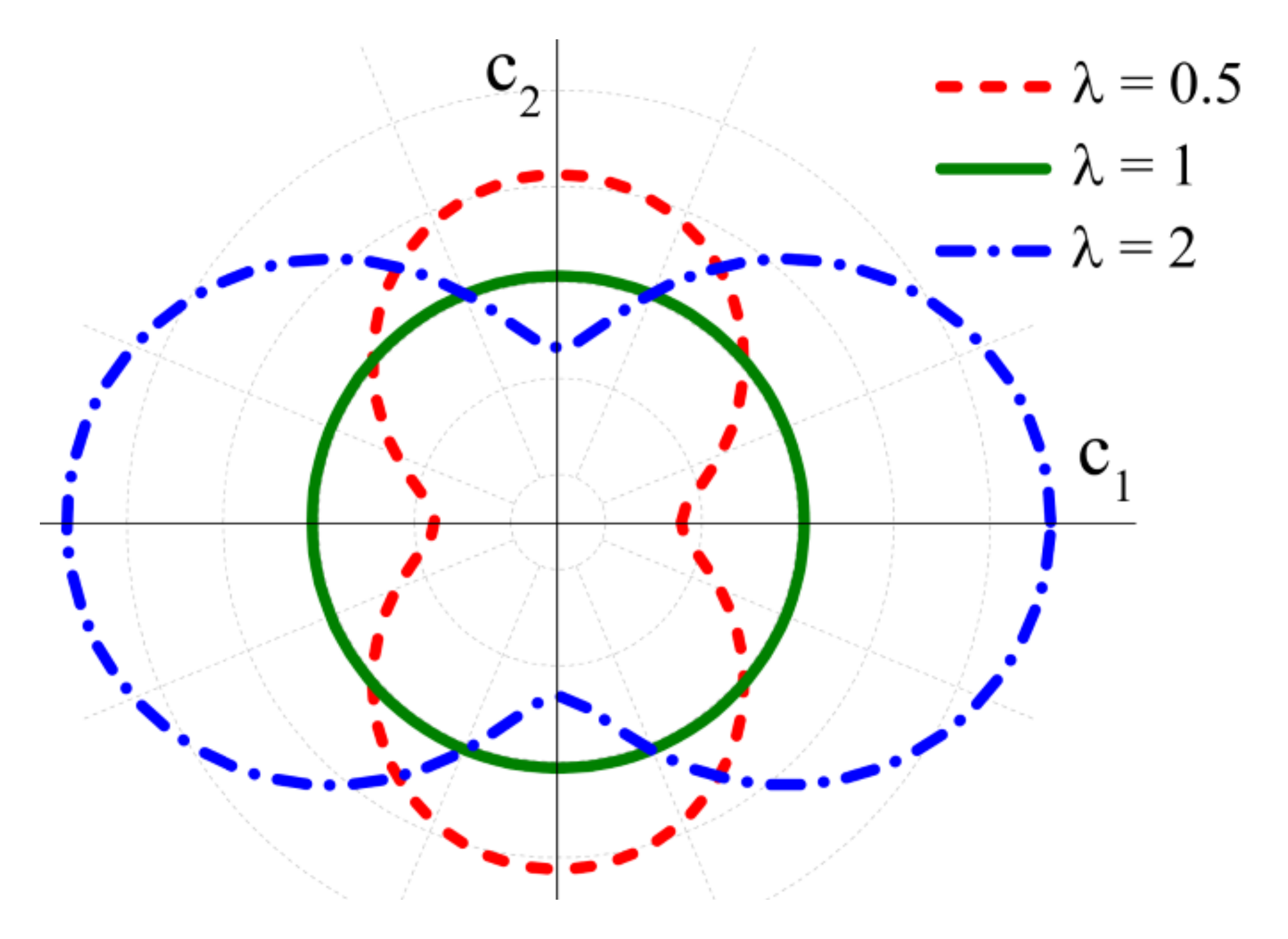}
 \caption{\footnotesize Polar diagram of the phase velocities for the shear waves ($K/\mu\gg 1)$.}\label{sw300}
 \end{figure}

 For nearly incompressible materials $(K/\mu\gg 1$ and $\tilde{\lambda}\simeq\lambda^{-1/2})$ the dependence of the pressure wave velocity on the direction of propagation and initial stress state is relatively weak. However, the velocities of shear waves depend strongly on the direction of propagation and the deformation state (FIG.~\ref{sw300}). In particular, we observe prominent minima and maxima of the velocities when waves propagate along the principal directions. Here and thereafter the velocities are normalized by the value in the undeformed state. 

\begin{figure}[h]
    \includegraphics[scale=0.25]{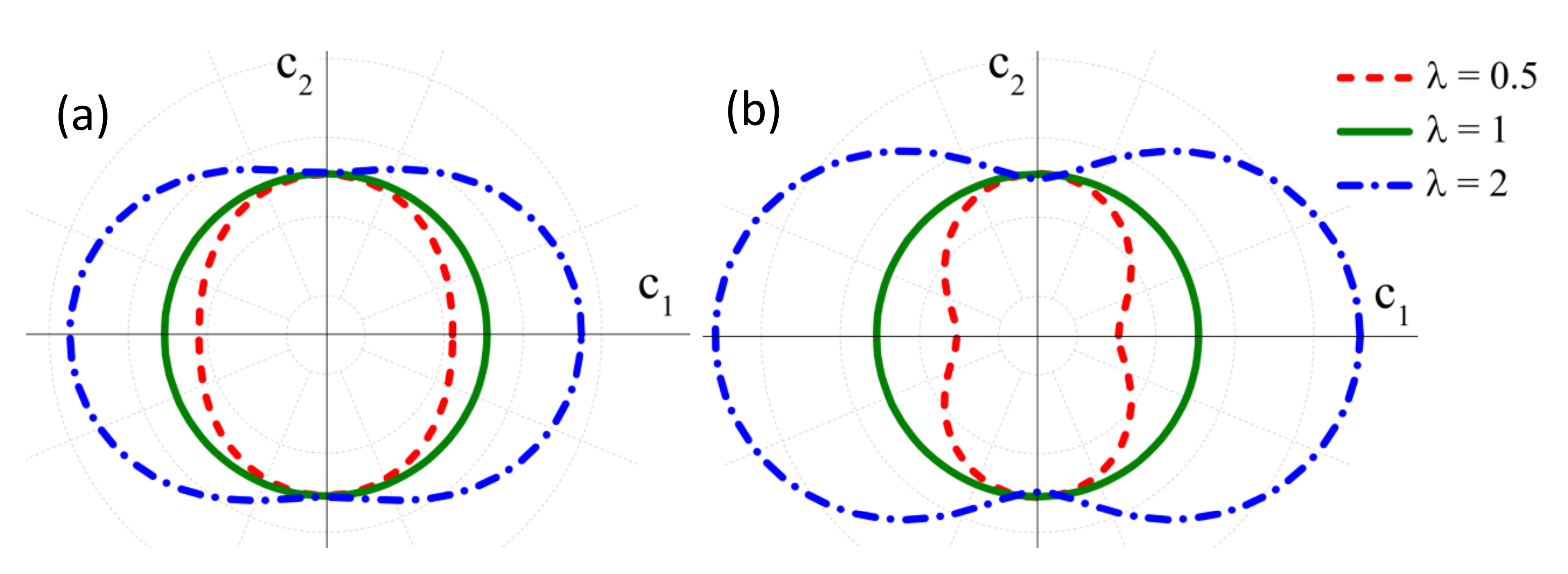}
   \caption{\footnotesize Polar diagrams of the phase velocities for the pressure (a) and shear (b) waves, $K/\mu=0.7$.}\label{k07m}
\end{figure}

 In contrast to nearly incompressible materials, for highly compressible matter, the velocity of pressure wave depends strongly on the direction of propagation and initial stress state (see FIG.~\ref{k07m} (a)). More specifically the velocity of pressure wave increases when material is stretched and decreases when it is compressed. Remarkably, if $K/\mu$ varies in the vicinity of $2/3$ which corresponds to Poisson's ratio $\nu$ in the vicinity of $0$, then the velocities of pressure and shear waves are not influenced by the initial stress-strain state when wave vector $\n$ lies in the plane of orthotropy orthogonal to the axis of tension ($\e_1$), see FIG.~\ref{k07m}.
Propagation of shear waves in highly compressible media differs significantly from the one in nearly incompressible materials. In particular, under compression the velocities of shear waves decrease in any direction if $K/\mu<1$ (FIG.~\ref{k07m} (b)), while it can either decrease or increase depending on the propagation direction in the nearly incompressible case (see FIG.~\ref{sw300}).

\begin{figure}[h]
   \includegraphics[scale=0.25]{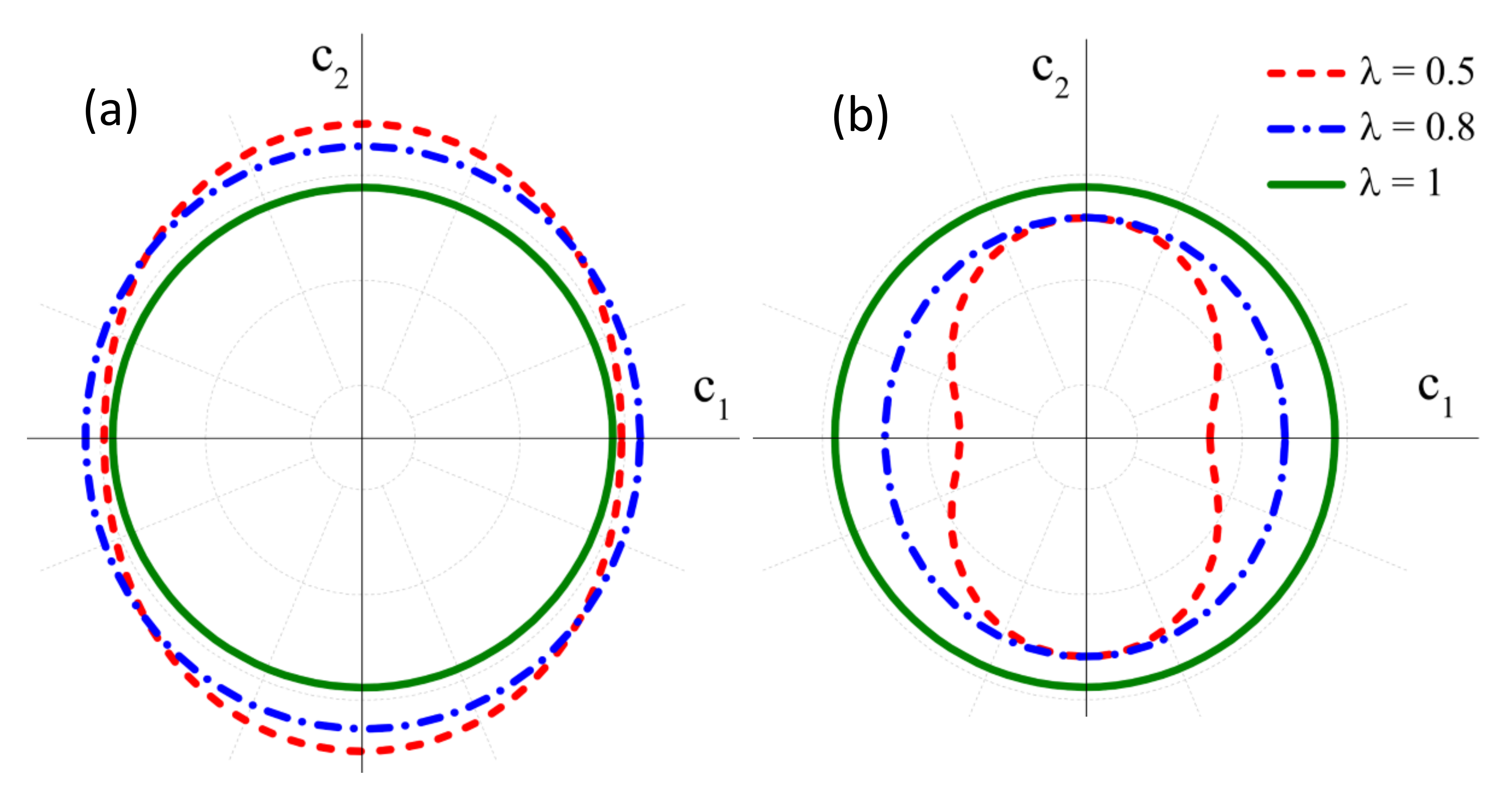}
  \caption{\footnotesize Polar diagrams of the phase velocities for the pressure (a) and shear (b) waves, $K/\mu=-0.3$.}\label{nk03m}
  \end{figure}

Next, we consider the auxetic materials with the negative bulk modulus. Yet can  such materials exist? Would they be stable? How to construct them? These questions have recently arisen in many papers\cite{Gasparetal2003jap,wanglakes2005,Chen2009,bertoldi&etal2010jam,bertoldi2013,Wang&etal2014prl,Still&etal2013prb,Beltramo&etal2014prl}  and still the topic is open for discussion. Wang and Lakes in their article\cite{wanglakes2005} report that bulk modulus can be varied within the range $-\frac{4\mu}{3}<K<\infty$. Based on this and our previous estimations, we examine the material behaviour when $-1/3<K/\mu<0$. An example for an auxetic material with $K/\mu=-0.3$ is shown in FIG.~\ref{nk03m} for pressure (a) and shear phase velocities (b).
We observe that the velocity of the pressure wave increases in any direction of propagation $\n$ when the material undergoes compression (FIG.~\ref{nk03m} (a)) in contrast to the case of positive bulk modulus (see FIG.~\ref{k07m} (a)). Moreover, the velocity of the pressure wave increases significantly and has maximum when wave vector $\n$ lies in the plane orthogonal to the axis of compression. Secondly, velocities of shear waves (FIG.~\ref{nk03m} (b)) decrease considerably for all directions of propagations $\n$ in contrast to the effect of deformation observed in the nearly incompressible materials (FIG.~\ref{sw300}).

The explicit expressions for corresponding angular frequencies are
\begin{gather}
  \omega_{pw}=\sqrt{\frac{(K-\frac{2\mu}{3})J^2k^2+\mu (k^2+\kk\cdot\B\cdot\kk)}{\rho_0}}\label{wpw}\\
    \text{and}\quad\omega_{sw}=\sqrt{\frac{\mu (\kk\cdot\B\cdot\kk)}{\rho_0}},\label{wsw}
\end{gather}
where $\kk$ is wave vector, $k=|\kk|$ is wave number and $\rho_0$ is the density of the undeformed material. Note that the wave vector can be also mapped into the undeformed configuration as $\kk^{0}=J^{-1}\Fder^{T}\kk$.

The transmission velocity of a wave packet or the group velocity can be found by the use of the formula $\mathbf{v}_{g}=\nabla_\kk\omega\ \Leftrightarrow\ \delta\omega=\delta\kk\cdot\mathbf{v}_{g}$.
 Thus, expressions \eqref{wpw} and \eqref{wsw} yield
\begin{equation}
\mathbf{v}_{g}^{pw}=\frac{\mu(\kk+\B\cdot\kk)+(K-\frac{2\mu}{3})J^2\kk}{\sqrt{\rho_0\left(\mu(k^2+\kk\cdot\B\cdot\kk)+(K-\frac{2\mu}{3})J^2k^2\right)}}\label{vgrpw}
\end{equation}
and
\begin{equation}
\mathbf{v}_{g}^{sw}=\B\cdot\kk\sqrt{\frac{\mu}{\rho_0(\kk\cdot\B\cdot\kk)}}\label{vgrsw}
\end{equation}

\begin{figure}[h]
     \includegraphics[scale=0.3]{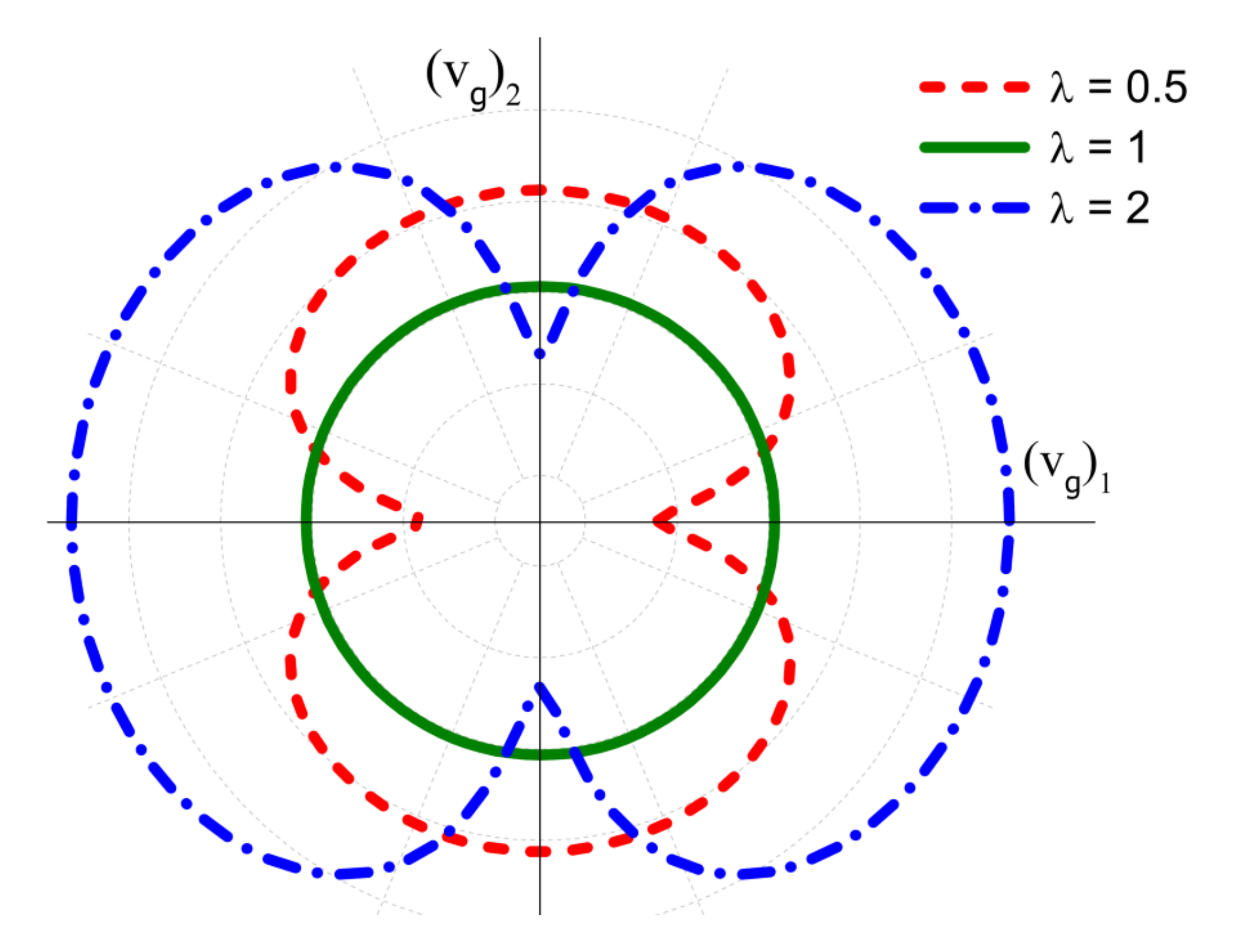}
    \caption{\footnotesize Polar diagram of energy curves for the shear waves $(K/\mu\gg 1)$.}\label{encur300}
   \end{figure}

To illustrate expressions $\eqref{vgrpw}$ and $\eqref{vgrsw}$ we plot the energy curves in FIG.~\ref{encur300}. Usually for homogeneous isotropic materials the velocity and energy curves coincide\cite{Auld1990,nayfeh1995}. However, in case of the nearly incompressible pre-stressed homogeneous isotropic material we find that velocities of energy and waves propagation coincide only for principal directions and when wave vector $\n$ lies in the plane orthogonal to the axis of tension. For other directions of propagation $\n$ they are significantly different (compare FIG.~\ref{encur300} and FIG.~\ref{sw300}).

 Next we examine the influence of deformation on wave propagation in materials characterized by strong stiffening effects. In particular, we make use of the Gent material model $\eqref{Gent}$; the corresponding acoustic tensor takes the form
  \begin{equation}\label{acGent}
  \begin{split}
  \mathbf{Q(n)}=\frac{\mu}{J}\left(1+\left(\frac{K}{\mu}-\frac{2}{3}-\frac{2}{J_m}\right)J^2\right)\n\otimes\n\\
  +\frac{\mu J_m}{J\xi^2}\left(\xi(\n\cdot\B\cdot\n)\I+2\n\otimes\n:\B\otimes\B^{(1324)}\right),
  \end{split}
  \end{equation}
  where $\xi=3+J_m-I_1$ and $\B\otimes\B^{(1324)}$ is isomer\cite{ryzhak93jmps} of tensor of fourth order, i.e. $\dd_1\otimes\dd_2\otimes\dd_3\otimes\dd_4^{(ijks)}=\dd_i\otimes\dd_j\otimes\dd_k\otimes\dd_s$, where (ijks) is a permutation of the set (1234).
One can see from $\eqref{acGent}$ that in general case the waves are neither pure longitudinal nor pure transverse in contrast to the waves prorogating in neo-Hookean materials. However, if wave vector $\n$ lies in one of the planes of orthotropy then we always have one pure transverse (shear) wave with the velocity
  \begin{equation}\label{cswGent}
	c_{sw}=\sqrt{a_2 J_m/(\rho\xi)}.
  \end{equation}
Note that expression $\eqref{cswGent}$ is general and valid for any finite deformation $\Fder$ and compressibility $K/\mu$.

\begin{figure}[h]
      \includegraphics[scale=0.3]{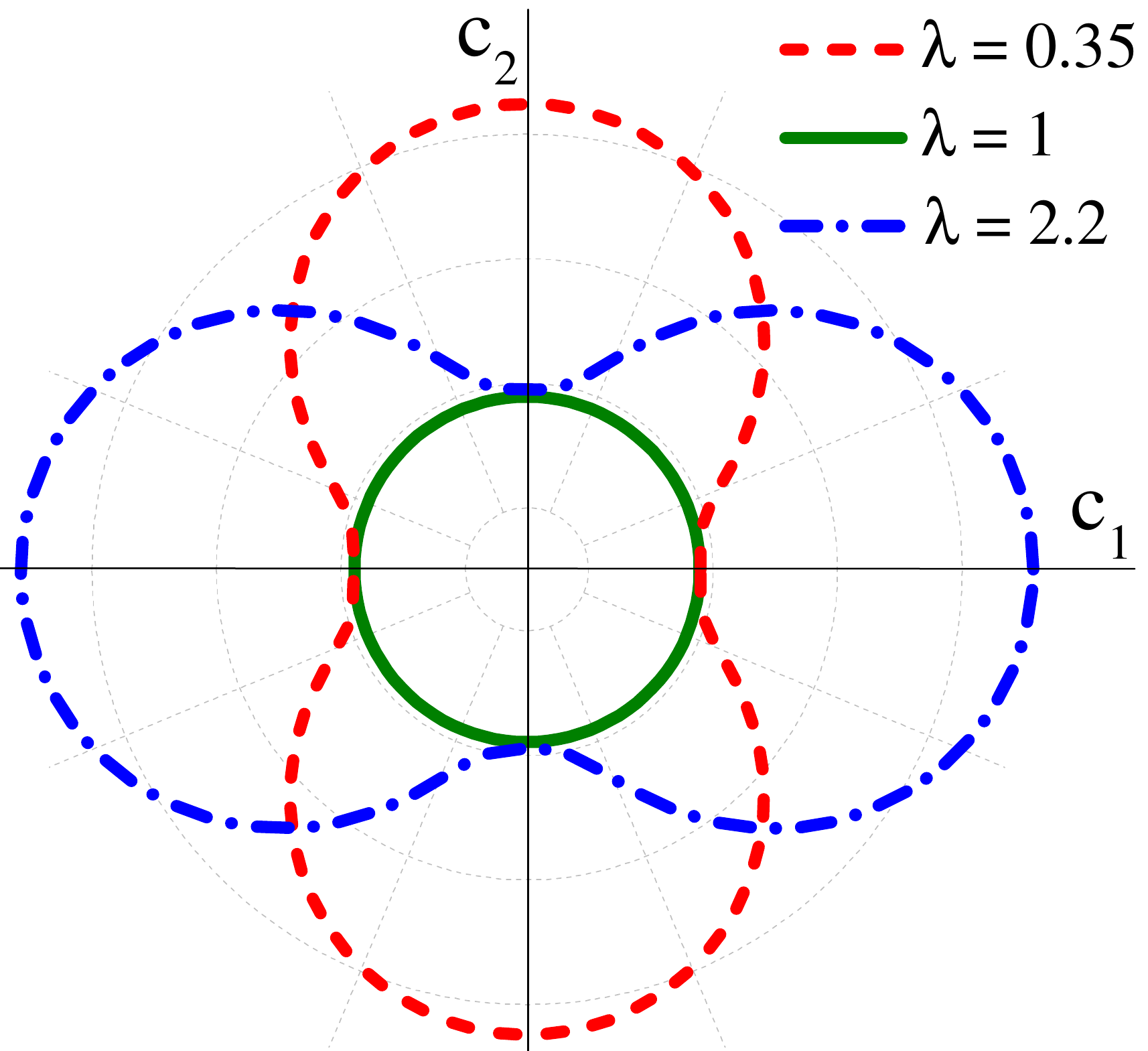}
      \caption{\footnotesize Polar diagram of the phase velocity for the quasi-pressure waves in Gent material ($J_m=3, K/\mu =300)$.}\label{GentPW}
\end{figure}

Similarly to the neo-Hookean results, for waves propagating along principal directions in Gent material, we have only pure modes. The corresponding expressions of velocities can be derived directly from $\eqref{acGent}$, however they are rather tedious. For waves propagating along the direction of loading ($\n=\e_1$) in nearly incompressible matter the velocities of longitudinal (pressure) and transverse (shear) waves reduce to
  \begin{equation}\label{cpwGent}
c_{pw}=\sqrt{\frac{K+\left(\frac{1}{3}-\frac{2}{J_m}+\frac{J_m\lambda\left(\lambda^3+(3+J_m)\lambda-2\right)}{\xi^2}\right)\mu}{\rho}}
  \end{equation}
 \begin{equation}\label{Gentcsw}
    \text{and}\quad c_{sw}=\lambda\sqrt{J_m\mu/(\rho\xi)}.
\end{equation}
It is easy to see from $\eqref{cpwGent}$ that for nearly incompressible materials the dependence of the longitudinal wave velocity is relatively weak unless extreme levels of deformations are attained. At these extreme deformations, the stiffening effect manifests in a significant increase of the effective shear modulus, which becomes comparable with the bulk modulus. We illustrate this dependence of the longitudinal wave velocity on the direction of propagation in FIG.~\ref{GentPW}.
We observe a dramatic change in the velocity profile of the longitudinal wave in extremely deformed materials. Moreover, the velocity of longitudinal wave increases in both compressed and stretched materials.

\begin{figure}[h]
      \includegraphics[scale=0.25]{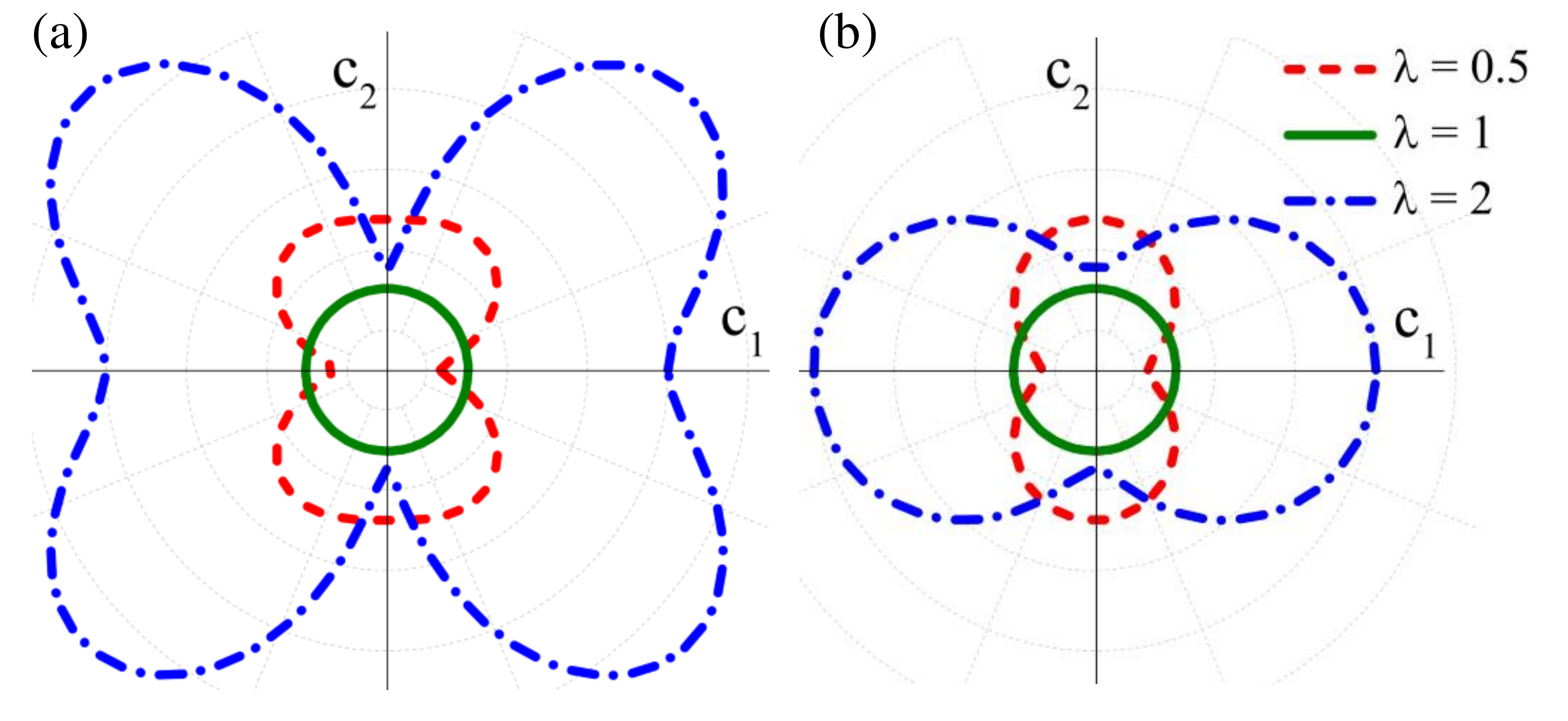}
      \caption{\footnotesize Polar diagrams of the phase velocities for the quasi-shear (a) and pure shear (b) waves in Gent material ($J_m=3, K/\mu\gg 1)$.}\label{GentSW}
\end{figure}

For the transverse waves, we find that the velocities depend strongly on the deformation and the stiffening parameter $\eqref{Gentcsw}$.
Examples of the dependence of the shear waves propagation on deformation for Gent material with $J_m=3$ and $K/\mu\gg 1$ are  shown on FIG.~\ref{GentSW}. The results clearly indicate the significant effect of the deformation induced stiffening (compare FIG.~\ref{sw300} and FIG.~\ref{GentSW}). We observe that the dependence of the shear wave velocity on deformation and prorogation direction increases when the locking parameter $J_m$ decreases (which corresponds to earlier stiffening of the material with deformation). Comparing the quasi shear wave velocity profiles with the pure shear ones, we observe that the velocity of quasi-shear wave has maxima for the non-principal directions (see FIG.~\ref{GentSW} (a)).

We derived the explicit dispersion relations for finitely deformed materials, and demonstrated the significant role of the deformation on the elastic wave prorogation on examples of both nearly incompressible and extreme auxetic materials. These findings may guide further design of mechanotunable acoustic metamaterials and phononic crystals with a large range of constituent properties. The local distribution of strain field in these engineered materials can be used to induce regions with extreme variation of the local phononic properties to give rise to various acoustic effects. This opens a very rich and broad research avenue for designing tunable acoustic/phononic metamaterials.

%

\nocite{*}
\bibliography{AuxeticsAPL}

\begin{thebibliography}{32}%
\makeatletter
\providecommand \@ifxundefined [1]{%
 \@ifx{#1\undefined}
}%
\providecommand \@ifnum [1]{%
 \ifnum #1\expandafter \@firstoftwo
 \else \expandafter \@secondoftwo
 \fi
}%
\providecommand \@ifx [1]{%
 \ifx #1\expandafter \@firstoftwo
 \else \expandafter \@secondoftwo
 \fi
}%
\providecommand \natexlab [1]{#1}%
\providecommand \enquote  [1]{``#1''}%
\providecommand \bibnamefont  [1]{#1}%
\providecommand \bibfnamefont [1]{#1}%
\providecommand \citenamefont [1]{#1}%
\providecommand \href@noop [0]{\@secondoftwo}%
\providecommand \href [0]{\begingroup \@sanitize@url \@href}%
\providecommand \@href[1]{\@@startlink{#1}\@@href}%
\providecommand \@@href[1]{\endgroup#1\@@endlink}%
\providecommand \@sanitize@url [0]{\catcode `\\12\catcode `\$12\catcode
  `\&12\catcode `\#12\catcode `\^12\catcode `\_12\catcode `\%12\relax}%
\providecommand \@@startlink[1]{}%
\providecommand \@@endlink[0]{}%
\providecommand \url  [0]{\begingroup\@sanitize@url \@url }%
\providecommand \@url [1]{\endgroup\@href {#1}{\urlprefix }}%
\providecommand \urlprefix  [0]{URL }%
\providecommand \Eprint [0]{\href }%
\providecommand \doibase [0]{http://dx.doi.org/}%
\providecommand \selectlanguage [0]{\@gobble}%
\providecommand \bibinfo  [0]{\@secondoftwo}%
\providecommand \bibfield  [0]{\@secondoftwo}%
\providecommand \translation [1]{[#1]}%
\providecommand \BibitemOpen [0]{}%
\providecommand \bibitemStop [0]{}%
\providecommand \bibitemNoStop [0]{.\EOS\space}%
\providecommand \EOS [0]{\spacefactor3000\relax}%
\providecommand \BibitemShut  [1]{\csname bibitem#1\endcsname}%
\let\auto@bib@innerbib\@empty
\bibitem [{\citenamefont {Biot}(1940)}]{Biot1940}%
  \BibitemOpen
  \bibfield  {author} {\bibinfo {author} {\bibfnamefont {M.}~\bibnamefont
  {Biot}},\ }\bibfield  {title} {\enquote {\bibinfo {title} {The influence of
  initial stress on elastic waves},}\ }\href@noop {} {\bibfield  {journal}
  {\bibinfo  {journal} {J. Appl. Phys.}\ }\textbf {\bibinfo {volume} {11}},\
  \bibinfo {pages} {522} (\bibinfo {year} {1940})}\BibitemShut {NoStop}%
\bibitem [{\citenamefont {Truesdell}\ and\ \citenamefont
  {Noll}(1965)}]{Truesdell1965}%
  \BibitemOpen
  \bibfield  {author} {\bibinfo {author} {\bibfnamefont {C.}~\bibnamefont
  {Truesdell}}\ and\ \bibinfo {author} {\bibfnamefont {W.}~\bibnamefont
  {Noll}},\ }\href@noop {} {\emph {\bibinfo {title} {The Non-Linear Field
  Theories of Mechanics}}},\ \bibinfo {edition} {3rd}\ ed.,\ edited by\
  \bibinfo {editor} {\bibfnamefont {S.~S.}\ \bibnamefont {Antman}}\ (\bibinfo
  {publisher} {Springer},\ \bibinfo {year} {1965})\BibitemShut {NoStop}%
\bibitem [{\citenamefont {Norris}(1983)}]{Norris1983}%
  \BibitemOpen
  \bibfield  {author} {\bibinfo {author} {\bibfnamefont {A.}~\bibnamefont
  {Norris}},\ }\bibfield  {title} {\enquote {\bibinfo {title} {Propagation of
  plane waves in a pre-stressed elastic medium},}\ }\href@noop {} {\bibfield
  {journal} {\bibinfo  {journal} {J. Acoust. Soc. Am}\ }\textbf {\bibinfo
  {volume} {74}},\ \bibinfo {pages} {1642} (\bibinfo {year}
  {1983})}\BibitemShut {NoStop}%
\bibitem [{\citenamefont {Ogden}(1984)}]{Ogden1984}%
  \BibitemOpen
  \bibfield  {author} {\bibinfo {author} {\bibfnamefont {R.}~\bibnamefont
  {Ogden}},\ }\href@noop {} {\emph {\bibinfo {title} {Non-Linear elastic
  deformations}}}\ (\bibinfo  {publisher} {Dover Publications},\ \bibinfo
  {year} {1984})\BibitemShut {NoStop}%
\bibitem [{\citenamefont {Kushwaha}\ \emph {et~al.}(1993)\citenamefont
  {Kushwaha}, \citenamefont {Halevi}, \citenamefont {Dobrzynski},\ and\
  \citenamefont {Djafari-Rouhani}}]{Kushwahaetal93prl}%
  \BibitemOpen
  \bibfield  {author} {\bibinfo {author} {\bibfnamefont {M.~S.}\ \bibnamefont
  {Kushwaha}}, \bibinfo {author} {\bibfnamefont {P.}~\bibnamefont {Halevi}},
  \bibinfo {author} {\bibfnamefont {L.}~\bibnamefont {Dobrzynski}}, \ and\
  \bibinfo {author} {\bibfnamefont {B.}~\bibnamefont {Djafari-Rouhani}},\
  }\bibfield  {title} {\enquote {\bibinfo {title} {Acoustic band structure of
  periodic elastic composites},}\ }\href@noop {} {\bibfield  {journal}
  {\bibinfo  {journal} {Phys. Rev. Lett.}\ }\textbf {\bibinfo {volume} {71}},\
  \bibinfo {pages} {2022} (\bibinfo {year} {1993})}\BibitemShut {NoStop}%
\bibitem [{\citenamefont {Ignatovich}\ and\ \citenamefont
  {Phan}(2009)}]{Ignatovich2009}%
  \BibitemOpen
  \bibfield  {author} {\bibinfo {author} {\bibfnamefont {V.~K.}\ \bibnamefont
  {Ignatovich}}\ and\ \bibinfo {author} {\bibfnamefont {L.~T.~N.}\ \bibnamefont
  {Phan}},\ }\bibfield  {title} {\enquote {\bibinfo {title} {Those wonderful
  elastic waves},}\ }\href@noop {} {\bibfield  {journal} {\bibinfo  {journal}
  {Am. J. Phys.}\ }\textbf {\bibinfo {volume} {77}},\ \bibinfo {pages} {1162}
  (\bibinfo {year} {2009})}\BibitemShut {NoStop}%
\bibitem [{\citenamefont {Ogden}\ and\ \citenamefont
  {Sighn}(2011)}]{OgdenSighn2011jmms}%
  \BibitemOpen
  \bibfield  {author} {\bibinfo {author} {\bibfnamefont {R.}~\bibnamefont
  {Ogden}}\ and\ \bibinfo {author} {\bibfnamefont {B.}~\bibnamefont {Sighn}},\
  }\bibfield  {title} {\enquote {\bibinfo {title} {Propagation of waves in an
  incompressible transversally isotripic elastic solid with initial stress:
  Biot revisited},}\ }\href@noop {} {\bibfield  {journal} {\bibinfo  {journal}
  {Journal of Mechanics of Materials and Structures}\ }\textbf {\bibinfo
  {volume} {6}},\ \bibinfo {pages} {453--477} (\bibinfo {year}
  {2011})}\BibitemShut {NoStop}%
\bibitem [{\citenamefont {Shams}, \citenamefont {Destrade},\ and\ \citenamefont
  {Ogden}(2011)}]{Shamsetal2011}%
  \BibitemOpen
  \bibfield  {author} {\bibinfo {author} {\bibfnamefont {M.}~\bibnamefont
  {Shams}}, \bibinfo {author} {\bibfnamefont {M.}~\bibnamefont {Destrade}}, \
  and\ \bibinfo {author} {\bibfnamefont {R.}~\bibnamefont {Ogden}},\ }\bibfield
   {title} {\enquote {\bibinfo {title} {Initial stresses in elastic solids:
  Constitutive laws and acoustoelasticity},}\ }\href@noop {} {\bibfield
  {journal} {\bibinfo  {journal} {Wave Motion}\ }\textbf {\bibinfo {volume}
  {48}},\ \bibinfo {pages} {552--567} (\bibinfo {year} {2011})}\BibitemShut
  {NoStop}%
\bibitem [{\citenamefont {Parnell}, \citenamefont {Norris},\ and\ \citenamefont
  {Shearer}(2012)}]{Parnell2012}%
  \BibitemOpen
  \bibfield  {author} {\bibinfo {author} {\bibfnamefont {W.}~\bibnamefont
  {Parnell}}, \bibinfo {author} {\bibfnamefont {A.~N.}\ \bibnamefont {Norris}},
  \ and\ \bibinfo {author} {\bibfnamefont {T.}~\bibnamefont {Shearer}},\
  }\bibfield  {title} {\enquote {\bibinfo {title} {Employing pre-stress to
  generate finite cloaks for antiplane elastic waves},}\ }\href@noop {}
  {\bibfield  {journal} {\bibinfo  {journal} {Appl. Phys. Lett.}\ }\textbf
  {\bibinfo {volume} {100}},\ \bibinfo {pages} {171907} (\bibinfo {year}
  {2012})}\BibitemShut {NoStop}%
\bibitem [{\citenamefont {Destrade}\ and\ \citenamefont
  {Ogden}(2013)}]{DestradeOgden13jam}%
  \BibitemOpen
  \bibfield  {author} {\bibinfo {author} {\bibfnamefont {M.}~\bibnamefont
  {Destrade}}\ and\ \bibinfo {author} {\bibfnamefont {R.~W.}\ \bibnamefont
  {Ogden}},\ }\bibfield  {title} {\enquote {\bibinfo {title} {On
  stress-dependent elastic moduli and wave speeds},}\ }\href@noop {} {\bibfield
   {journal} {\bibinfo  {journal} {J. of Appl. Math.}\ }\textbf {\bibinfo
  {volume} {78}},\ \bibinfo {pages} {965--997} (\bibinfo {year}
  {2013})}\BibitemShut {NoStop}%
\bibitem [{\citenamefont {Hussein}, \citenamefont {Leamy},\ and\ \citenamefont
  {Ruzzene}(2014)}]{Hussein2014}%
  \BibitemOpen
  \bibfield  {author} {\bibinfo {author} {\bibfnamefont {M.}~\bibnamefont
  {Hussein}}, \bibinfo {author} {\bibfnamefont {M.}~\bibnamefont {Leamy}}, \
  and\ \bibinfo {author} {\bibfnamefont {M.}~\bibnamefont {Ruzzene}},\
  }\bibfield  {title} {\enquote {\bibinfo {title} {Dynamics of phononic
  materials and structures: Historical origins, recent progress and future
  outlook},}\ }\href@noop {} {\bibfield  {journal} {\bibinfo  {journal}
  {Applied Mechanics Reviews}\ }\textbf {\bibinfo {volume} {66}},\ \bibinfo
  {pages} {040802} (\bibinfo {year} {2014})}\BibitemShut {NoStop}%
\bibitem [{\citenamefont {Wang}\ \emph {et~al.}(2014)\citenamefont {Wang},
  \citenamefont {Casadei}, \citenamefont {Shan}, \citenamefont {Weaver},\ and\
  \citenamefont {Bertoldi}}]{Wang&etal2014prl}%
  \BibitemOpen
  \bibfield  {author} {\bibinfo {author} {\bibfnamefont {P.}~\bibnamefont
  {Wang}}, \bibinfo {author} {\bibfnamefont {F.}~\bibnamefont {Casadei}},
  \bibinfo {author} {\bibfnamefont {S.}~\bibnamefont {Shan}}, \bibinfo {author}
  {\bibfnamefont {J.~C.}\ \bibnamefont {Weaver}}, \ and\ \bibinfo {author}
  {\bibfnamefont {K.}~\bibnamefont {Bertoldi}},\ }\bibfield  {title} {\enquote
  {\bibinfo {title} {Harnessing buckling to design tunable locally resonant
  acoustic metamaterials},}\ }\href@noop {} {\bibfield  {journal} {\bibinfo
  {journal} {Phys. Rev. Lett.}\ }\textbf {\bibinfo {volume} {113}},\ \bibinfo
  {pages} {014301} (\bibinfo {year} {2014})}\BibitemShut {NoStop}%
\bibitem [{\citenamefont {Zhang}, \citenamefont {Xia},\ and\ \citenamefont
  {Fang}(2011)}]{Fang2011prl}%
  \BibitemOpen
  \bibfield  {author} {\bibinfo {author} {\bibfnamefont {S.}~\bibnamefont
  {Zhang}}, \bibinfo {author} {\bibfnamefont {C.}~\bibnamefont {Xia}}, \ and\
  \bibinfo {author} {\bibfnamefont {N.}~\bibnamefont {Fang}},\ }\bibfield
  {title} {\enquote {\bibinfo {title} {Broadband acoustic cloak for ultrasound
  waves},}\ }\href@noop {} {\bibfield  {journal} {\bibinfo  {journal} {Phys.
  Rev. Lett.}\ }\textbf {\bibinfo {volume} {106}},\ \bibinfo {pages} {024301}
  (\bibinfo {year} {2011})}\BibitemShut {NoStop}%
\bibitem [{\citenamefont {Bertoldi}\ and\ \citenamefont
  {Boyce}(2008)}]{BertoldiBoyce2008prb}%
  \BibitemOpen
  \bibfield  {author} {\bibinfo {author} {\bibfnamefont {K.}~\bibnamefont
  {Bertoldi}}\ and\ \bibinfo {author} {\bibfnamefont {M.~C.}\ \bibnamefont
  {Boyce}},\ }\bibfield  {title} {\enquote {\bibinfo {title} {Wave propagation
  and instabilities in monolithic and periodically structured elastomeric
  materials undergoing large deformations},}\ }\href@noop {} {\bibfield
  {journal} {\bibinfo  {journal} {Phys. Rev. B}\ }\textbf {\bibinfo {volume}
  {78}},\ \bibinfo {pages} {184107} (\bibinfo {year} {2008})}\BibitemShut
  {NoStop}%
\bibitem [{\citenamefont {Rudykh}\ and\ \citenamefont
  {Boyce}(2014)}]{RudykhBoyce2014prl}%
  \BibitemOpen
  \bibfield  {author} {\bibinfo {author} {\bibfnamefont {S.}~\bibnamefont
  {Rudykh}}\ and\ \bibinfo {author} {\bibfnamefont {M.}~\bibnamefont {Boyce}},\
  }\bibfield  {title} {\enquote {\bibinfo {title} {Transforming wave
  propagation in layered media via instability-induced interfacial
  wrinkling},}\ }\href@noop {} {\bibfield  {journal} {\bibinfo  {journal}
  {Phys. Rev. Lett.}\ }\textbf {\bibinfo {volume} {112}},\ \bibinfo {pages}
  {034301} (\bibinfo {year} {2014})}\BibitemShut {NoStop}%
\bibitem [{\citenamefont {Li}\ \emph {et~al.}(2013)\citenamefont {Li},
  \citenamefont {Kaynia}, \citenamefont {Rudykh},\ and\ \citenamefont
  {Boyce}}]{Li13etal}%
  \BibitemOpen
  \bibfield  {author} {\bibinfo {author} {\bibfnamefont {Y.}~\bibnamefont
  {Li}}, \bibinfo {author} {\bibfnamefont {N.}~\bibnamefont {Kaynia}}, \bibinfo
  {author} {\bibfnamefont {S.}~\bibnamefont {Rudykh}}, \ and\ \bibinfo {author}
  {\bibfnamefont {M.}~\bibnamefont {Boyce}},\ }\bibfield  {title} {\enquote
  {\bibinfo {title} {Wrinkling of interfacial layers in stratified
  composites},}\ }\href@noop {} {\bibfield  {journal} {\bibinfo  {journal}
  {Advanced Engineering Materials}\ }\textbf {\bibinfo {volume} {15}},\
  \bibinfo {pages} {921--926} (\bibinfo {year} {2013})}\BibitemShut {NoStop}%
\bibitem [{\citenamefont {Lydon}, \citenamefont {Serra-Garcia},\ and\
  \citenamefont {Daraio}(2014)}]{Lydon&etal2014prl}%
  \BibitemOpen
  \bibfield  {author} {\bibinfo {author} {\bibfnamefont {J.}~\bibnamefont
  {Lydon}}, \bibinfo {author} {\bibfnamefont {M.}~\bibnamefont {Serra-Garcia}},
  \ and\ \bibinfo {author} {\bibfnamefont {C.}~\bibnamefont {Daraio}},\
  }\bibfield  {title} {\enquote {\bibinfo {title} {Local to extended
  transitions of resonant defect modes},}\ }\href@noop {} {\bibfield  {journal}
  {\bibinfo  {journal} {Phys. Rev. Lett.}\ }\textbf {\bibinfo {volume} {113}},\
  \bibinfo {pages} {185503} (\bibinfo {year} {2014})}\BibitemShut {NoStop}%
\bibitem [{\citenamefont {Williams}\ and\ \citenamefont
  {Lewis}(1982)}]{WILLIAMS1982}%
  \BibitemOpen
  \bibfield  {author} {\bibinfo {author} {\bibfnamefont {J.~L.}\ \bibnamefont
  {Williams}}\ and\ \bibinfo {author} {\bibfnamefont {J.~L.}\ \bibnamefont
  {Lewis}},\ }\bibfield  {title} {\enquote {\bibinfo {title} {Properties and an
  anisotropic model of cancellous bone from the proximal tibial epiphysis},}\
  }\href@noop {} {\bibfield  {journal} {\bibinfo  {journal} {Trans.. ASME, J.
  Biomech. Eng.}\ }\textbf {\bibinfo {volume} {104}},\ \bibinfo {pages} {5--56}
  (\bibinfo {year} {1982})}\BibitemShut {NoStop}%
\bibitem [{\citenamefont {Evans}\ and\ \citenamefont
  {Alderson}(2000)}]{Evans2000}%
  \BibitemOpen
  \bibfield  {author} {\bibinfo {author} {\bibfnamefont {K.~E.}\ \bibnamefont
  {Evans}}\ and\ \bibinfo {author} {\bibfnamefont {K.~L.}\ \bibnamefont
  {Alderson}},\ }\bibfield  {title} {\enquote {\bibinfo {title} {Auxetic
  materials: the positive side of being negative},}\ }\href@noop {} {\bibfield
  {journal} {\bibinfo  {journal} {Engineering Science and Education Journal}\
  }\textbf {\bibinfo {volume} {9}},\ \bibinfo {pages} {148--154} (\bibinfo
  {year} {2000})}\BibitemShut {NoStop}%
\bibitem [{\citenamefont {Caddock}\ and\ \citenamefont
  {Evans}(1995)}]{Caddock1995}%
  \BibitemOpen
  \bibfield  {author} {\bibinfo {author} {\bibfnamefont {B.~D.}\ \bibnamefont
  {Caddock}}\ and\ \bibinfo {author} {\bibfnamefont {K.~E.}\ \bibnamefont
  {Evans}},\ }\bibfield  {title} {\enquote {\bibinfo {title} {Negative
  poisson's ratios and strain-dependent mechanical properties in arterial
  prostheses},}\ }\href@noop {} {\bibfield  {journal} {\bibinfo  {journal}
  {Biomaterials}\ }\textbf {\bibinfo {volume} {16}},\ \bibinfo {pages}
  {1109--1115} (\bibinfo {year} {1995})}\BibitemShut {NoStop}%
\bibitem [{\citenamefont {Babaee}\ \emph {et~al.}(2013)\citenamefont {Babaee},
  \citenamefont {Shim}, \citenamefont {Weaver}, \citenamefont {Chen},
  \citenamefont {Patel},\ and\ \citenamefont {Bertoldi}}]{bertoldi2013}%
  \BibitemOpen
  \bibfield  {author} {\bibinfo {author} {\bibfnamefont {S.}~\bibnamefont
  {Babaee}}, \bibinfo {author} {\bibfnamefont {J.}~\bibnamefont {Shim}},
  \bibinfo {author} {\bibfnamefont {J.}~\bibnamefont {Weaver}}, \bibinfo
  {author} {\bibfnamefont {E.}~\bibnamefont {Chen}}, \bibinfo {author}
  {\bibfnamefont {N.}~\bibnamefont {Patel}}, \ and\ \bibinfo {author}
  {\bibfnamefont {K.}~\bibnamefont {Bertoldi}},\ }\bibfield  {title} {\enquote
  {\bibinfo {title} {3d soft metamaterials with negative poisson's ratio},}\
  }\href@noop {} {\bibfield  {journal} {\bibinfo  {journal} {Adv. Mater.}\
  }\textbf {\bibinfo {volume} {25}},\ \bibinfo {pages} {5044--5049} (\bibinfo
  {year} {2013})}\BibitemShut {NoStop}%
\bibitem [{\citenamefont {Arruda}\ and\ \citenamefont
  {Boyce}(1993)}]{Arruda1993}%
  \BibitemOpen
  \bibfield  {author} {\bibinfo {author} {\bibfnamefont {E.}~\bibnamefont
  {Arruda}}\ and\ \bibinfo {author} {\bibfnamefont {M.}~\bibnamefont {Boyce}},\
  }\bibfield  {title} {\enquote {\bibinfo {title} {A three-dimensional
  constitutive model for the large stretch behavior of rubber elastic
  materials},}\ }\href@noop {} {\bibfield  {journal} {\bibinfo  {journal} {J.
  Mech. Phys. Solids}\ }\textbf {\bibinfo {volume} {41}},\ \bibinfo {pages}
  {389–412} (\bibinfo {year} {1993})}\BibitemShut {NoStop}%
\bibitem [{\citenamefont {Gent}(1996)}]{gent1996}%
  \BibitemOpen
  \bibfield  {author} {\bibinfo {author} {\bibfnamefont {A.~N.}\ \bibnamefont
  {Gent}},\ }\bibfield  {title} {\enquote {\bibinfo {title} {A new constitutive
  relation for rubber},}\ }\href@noop {} {\bibfield  {journal} {\bibinfo
  {journal} {Rubber Chemistry and Technology}\ }\textbf {\bibinfo {volume}
  {69}},\ \bibinfo {pages} {59--61} (\bibinfo {year} {1996})}\BibitemShut
  {NoStop}%
\bibitem [{\citenamefont {Ryzhak}(1993)}]{ryzhak93jmps}%
  \BibitemOpen
  \bibfield  {author} {\bibinfo {author} {\bibfnamefont {E.~I.}\ \bibnamefont
  {Ryzhak}},\ }\bibfield  {title} {\enquote {\bibinfo {title} {On stable
  deformation of "unstable" materials in a rigid triaxial testing mashine},}\
  }\href@noop {} {\bibfield  {journal} {\bibinfo  {journal} {J. Mech. Phys.
  Solids}\ }\textbf {\bibinfo {volume} {41}},\ \bibinfo {pages} {1345--1356}
  (\bibinfo {year} {1993})}\BibitemShut {NoStop}%
\bibitem [{\citenamefont {Wang}\ and\ \citenamefont
  {Lakes}(2005)}]{wanglakes2005}%
  \BibitemOpen
  \bibfield  {author} {\bibinfo {author} {\bibfnamefont {Y.}~\bibnamefont
  {Wang}}\ and\ \bibinfo {author} {\bibfnamefont {R.}~\bibnamefont {Lakes}},\
  }\bibfield  {title} {\enquote {\bibinfo {title} {Composites with inclusions
  of negative bulk modulus: Extreme damping and negative poisson's ratio},}\
  }\href@noop {} {\bibfield  {journal} {\bibinfo  {journal} {Journal of
  Composite Materials}\ }\textbf {\bibinfo {volume} {39}},\ \bibinfo {pages}
  {1645--1657} (\bibinfo {year} {2005})}\BibitemShut {NoStop}%
\bibitem [{\citenamefont {Gaspar}, \citenamefont {Smith},\ and\ \citenamefont
  {Evans}(2003)}]{Gasparetal2003jap}%
  \BibitemOpen
  \bibfield  {author} {\bibinfo {author} {\bibfnamefont {N.}~\bibnamefont
  {Gaspar}}, \bibinfo {author} {\bibfnamefont {C.~W.}\ \bibnamefont {Smith}}, \
  and\ \bibinfo {author} {\bibfnamefont {K.~E.}\ \bibnamefont {Evans}},\
  }\bibfield  {title} {\enquote {\bibinfo {title} {Effect of heterogeneity on
  the elastic properties of auxetic materials},}\ }\href@noop {} {\bibfield
  {journal} {\bibinfo  {journal} {J. Appl. Phys.}\ }\textbf {\bibinfo {volume}
  {94}},\ \bibinfo {pages} {6143} (\bibinfo {year} {2003})}\BibitemShut
  {NoStop}%
\bibitem [{\citenamefont {Chen}\ \emph {et~al.}(2009)\citenamefont {Chen},
  \citenamefont {Liu}, \citenamefont {Wang}, \citenamefont {Zhang},
  \citenamefont {Hu},\ and\ \citenamefont {Fan}}]{Chen2009}%
  \BibitemOpen
  \bibfield  {author} {\bibinfo {author} {\bibfnamefont {L.}~\bibnamefont
  {Chen}}, \bibinfo {author} {\bibfnamefont {C.}~\bibnamefont {Liu}}, \bibinfo
  {author} {\bibfnamefont {J.}~\bibnamefont {Wang}}, \bibinfo {author}
  {\bibfnamefont {W.}~\bibnamefont {Zhang}}, \bibinfo {author} {\bibfnamefont
  {C.}~\bibnamefont {Hu}}, \ and\ \bibinfo {author} {\bibfnamefont
  {S.}~\bibnamefont {Fan}},\ }\bibfield  {title} {\enquote {\bibinfo {title}
  {Auxetic materials with large negative poisson's ratios based on highly
  oriented carbon nanotube structures},}\ }\href@noop {} {\bibfield  {journal}
  {\bibinfo  {journal} {Appl. Phys. Lett.}\ }\textbf {\bibinfo {volume} {94}},\
  \bibinfo {pages} {253111} (\bibinfo {year} {2009})}\BibitemShut {NoStop}%
\bibitem [{\citenamefont {Bertoldi}\ \emph {et~al.}(2010)\citenamefont
  {Bertoldi}, \citenamefont {Reis}, \citenamefont {Willshaw},\ and\
  \citenamefont {Mullin}}]{bertoldi&etal2010jam}%
  \BibitemOpen
  \bibfield  {author} {\bibinfo {author} {\bibfnamefont {K.}~\bibnamefont
  {Bertoldi}}, \bibinfo {author} {\bibfnamefont {P.}~\bibnamefont {Reis}},
  \bibinfo {author} {\bibfnamefont {S.}~\bibnamefont {Willshaw}}, \ and\
  \bibinfo {author} {\bibfnamefont {T.}~\bibnamefont {Mullin}},\ }\bibfield
  {title} {\enquote {\bibinfo {title} {Negative poisson's ratio behavior
  induced by an elastic instability},}\ }\href@noop {} {\bibfield  {journal}
  {\bibinfo  {journal} {Adv. Mater}\ }\textbf {\bibinfo {volume} {22}},\
  \bibinfo {pages} {361--366} (\bibinfo {year} {2010})}\BibitemShut {NoStop}%
\bibitem [{\citenamefont {Still}\ \emph {et~al.}(2013)\citenamefont {Still},
  \citenamefont {Oudich}, \citenamefont {Auerhammer}, \citenamefont
  {Vlassopoulos}, \citenamefont {Djafari-Rouhani}, \citenamefont {Fytas},\ and\
  \citenamefont {Sheng}}]{Still&etal2013prb}%
  \BibitemOpen
  \bibfield  {author} {\bibinfo {author} {\bibfnamefont {T.}~\bibnamefont
  {Still}}, \bibinfo {author} {\bibfnamefont {M.}~\bibnamefont {Oudich}},
  \bibinfo {author} {\bibfnamefont {G.~K.}\ \bibnamefont {Auerhammer}},
  \bibinfo {author} {\bibfnamefont {D.}~\bibnamefont {Vlassopoulos}}, \bibinfo
  {author} {\bibfnamefont {B.}~\bibnamefont {Djafari-Rouhani}}, \bibinfo
  {author} {\bibfnamefont {G.}~\bibnamefont {Fytas}}, \ and\ \bibinfo {author}
  {\bibfnamefont {P.}~\bibnamefont {Sheng}},\ }\bibfield  {title} {\enquote
  {\bibinfo {title} {Soft silicone rubber in phononic structures: Correct
  elastic moduli},}\ }\href@noop {} {\bibfield  {journal} {\bibinfo  {journal}
  {Phys. Rev. B}\ }\textbf {\bibinfo {volume} {88}},\ \bibinfo {pages} {094102}
  (\bibinfo {year} {2013})}\BibitemShut {NoStop}%
\bibitem [{\citenamefont {Beltramo}\ \emph {et~al.}(2014)\citenamefont
  {Beltramo}, \citenamefont {Schneider}, \citenamefont {Fytas},\ and\
  \citenamefont {Furst}}]{Beltramo&etal2014prl}%
  \BibitemOpen
  \bibfield  {author} {\bibinfo {author} {\bibfnamefont {P.~J.}\ \bibnamefont
  {Beltramo}}, \bibinfo {author} {\bibfnamefont {D.}~\bibnamefont {Schneider}},
  \bibinfo {author} {\bibfnamefont {G.}~\bibnamefont {Fytas}}, \ and\ \bibinfo
  {author} {\bibfnamefont {E.~M.}\ \bibnamefont {Furst}},\ }\bibfield  {title}
  {\enquote {\bibinfo {title} {Anisotropic hypersonic phonon propagation in
  films of aligned ellipsoids},}\ }\href@noop {} {\bibfield  {journal}
  {\bibinfo  {journal} {Phys. Rev. Lett.}\ }\textbf {\bibinfo {volume} {113}},\
  \bibinfo {pages} {205503} (\bibinfo {year} {2014})}\BibitemShut {NoStop}%
\bibitem [{\citenamefont {Auld}(1990)}]{Auld1990}%
  \BibitemOpen
  \bibfield  {author} {\bibinfo {author} {\bibfnamefont {B.}~\bibnamefont
  {Auld}},\ }\href@noop {} {\emph {\bibinfo {title} {Acoustic fields and waves
  in solids}}}\ (\bibinfo  {publisher} {Krieger publishing company},\ \bibinfo
  {year} {1990})\BibitemShut {NoStop}%
\bibitem [{\citenamefont {Nayfeh}(1995)}]{nayfeh1995}%
  \BibitemOpen
  \bibfield  {author} {\bibinfo {author} {\bibfnamefont {A.}~\bibnamefont
  {Nayfeh}},\ }\href@noop {} {\emph {\bibinfo {title} {Wave propagation in
  layered anisotropic media with applications to composites}}}\ (\bibinfo
  {publisher} {Elsevier Science},\ \bibinfo {year} {1995})\BibitemShut
  {NoStop}%
\end{thebibliography}%

\end{document}